■   3523

# Modeling and Performance of Microwave and Millimeter-Wave Layered Waveguide Filters

**Ardavan Rahimian**
School of Electronic, Electrical and Computer Engineering, University of Birmingham
Edgbaston, Birmingham B15 2TT, UK
e-mail: rahimian@ieee.org

*Abstract*

*This paper presents novel designs, analysis, and performance of 4-pole and 8-pole microwave and millimeter-wave (MMW) waveguide filters for operation at X and Y frequency bands. The waveguide filters have been designed and analyzed based on the RF mode matching and coupled resonators design techniques employing layered technology. Thorough waveguide filters working at X-band and Y-band have been designed, analyzed, fabricated, and also tested along with the analysis of the output characteristics. Accurate designs of RF waveguides along with their filters based on the E-plane filter concept have been carried out with the ability of fitting into the layered technology in high frequency production techniques. The filters demonstrate the appropriateness in order to develop high-performance well-established designs for systems that are intended for the multi-layer microwave, millimeter- and sub-millimeter-waves devices and systems; with the potential employment in radar, satellite, and radio astronomy applications.*

*Keywords: coupled resonator, millimeter-wave device, RF filtering function, RF waveguide filter*

## 1. Introduction

The demands for extending the applications of RF waveguide devices and systems to the increasingly higher RF frequencies up to sub-millimeter-wave frequency bands have been driven by applications including astronomy, radars, and remote sensors [1]. This increased need for higher bandwidth systems, results significantly in microwave and millimeter-wave (MMW) communications become the crucial importance of modern systems. Hence, microwave filters are the important devices for RF/microwave communication systems. This two-port device that provides selectivity in high frequencies has numerous applications in satellite communications, and radio scanning systems. RF Filters play an important role to facilitate frequency agility in radars, satellite, and cellular communications. The aim of the RF filtering function stage is to select the desired frequency components and to reduce those bringing no information [2]. The filter performance reflects hence its ability to get the desired information without any degradation and to reject the interfering signals down to low levels, so that their existence does not degrade the overall system performance [3]. The microwave and MMW filters are a promising technology useful to prevent signals from interfering communications. Especially, it is required for recent trends in modern systems that need smart and also reconfigurable transceivers to meet multiple standards and functions. Both analytical and numerical RF methods for synthesis of coupling matrices corresponding to the RF and microwave filters have been extensively studied [4].

Modeling and design of these devices based on the computer aided design (CAD) of components has been employed to obtain the accurate physical dimensions with the prescribed specifications, and reduce the experimental debugging and tuning period after the possible manufacture of the RF devices. The employed techniques for designing these filters are based on the mode matching and coupling coefficients for devices arranged in a topology representing a two-port microwave network. Optimization techniques can further be employed in order to synthesize the coupling matrix by minimization of a scalar cost function. However, fabrication of waveguide components at high frequencies using traditional methods, such a metal milling or electrical discharge machining can be expensive and suffers from the lack of dimensional accuracy [5]. The advantage of these filters over the similar microstrip planar structures is their *Q*-factor that enables satisfactory stop-band attenuation. Such filters have low losses and slight coupling between resonators even more increases the *Q*-factor [6], making these waveguide filters suitable for narrow bandwidths [7]. In this contribution, the comprehensive designs of *E*-plane RF waveguide filters have been carried out for *X*-band and *Y*-band, and also their output characteristics are presented and analyzed. The RF filter designs include two *X*-band 4-pole band-pass microwave filters, and one *X*-band 8-pole band-pass microwave filter, which have been simulated, fabricated, and tested based on the mode matching and coupled resonators and coupled resonators, respectively. Their output characteristics have been analyzed in terms of scattering (*S*) parameters along with introducing a novel microwave layered structure. A *Y*-band 4-pole band-pass *E*-plane MMW filter based on the coupled resonators design method has also been accurately designed and simulated and its output RF system characteristics have been shown and analyzed in terms of the microwave scattering parameters in detail as well.

## 2. Waveguide *E*-Plane Filter Analysis

A waveguide is known as any structure that has the ability of propagating the energy along a pre-determined path in form of the electromagnetic (EM) wave and energy, and in order to couple elements together to channel the signal around. Also, the ideal waveguide performs this task without the loss of energy and distortion of electromagnetic wave; real waveguides are approximate to the ideal one [8]. RF waveguide devices have different structures dependent on bandwidth. Rectangular waveguides are utilized at high frequencies and prevent radiation and loss by the sidewalls. Layout of the simulated waveguide in the proposed *X*-band (WG16) and *Y*-band (WR3) have also been shown below in figure 1. According to the EM theory of boundary conditions, tangential component of the electric field *E* and normal component of the magnetic field *H* are both zero at the surface of a good conductor. Propagation is not possible for waves with wavelength above twice of the RF waveguide width ($\lambda_c = 2a$). The corresponding frequency is called the RF cut-off frequency and $\lambda_c$ is called the cut-off wavelength. The EM wave reflects back and forth between the walls and also does not flow down the guide for $\lambda = \lambda_c$ [8]. There are currents flowing in the walls and as the magnetic field is tangential at the surface of a conductor, the current flow is perpendicular to *H*. Current on the surface of the waveguide, has its highest value in middle of top and bottom surfaces and decreases down to zero as it gets closer to sidewalls. Hence, there is no current on waveguide sidewalls. Direct-coupled *E*-plane filters can be constructed by inserting a ladder-type metallic in the center of a waveguide, which offer the potential of realizing low cost mass-producible and low dissipation loss microwave filters [9]. The pure metal insert consists of the resonators and septa (coupling). When the wave enters the filter, it propagates through the resonator and hits the metal part. A part of the wave will be reflected back and resonates in the resonator. The other part will pass through the coupling and as the coupling length is wider, the EM coupling will be smaller in the RF waveguide device.

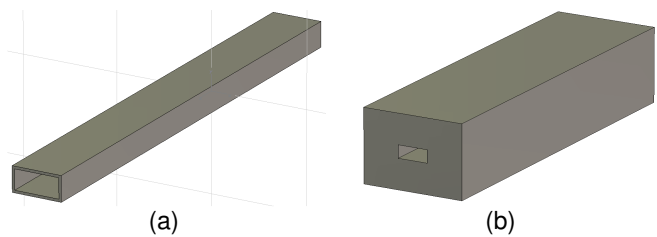

Figure 1. Layouts of the RF rectangular waveguides:
(a) *X*-band (WG16), cut-off frequency: 6.56 GHz, range (GHz): 8.2-12.4;
(b) *Y*-band (WR3), cut-off frequency: 196.71 GHz, range (GHz): 220-325.

### 2.1. Method 1: Mode Matching Technique

The RF method is based on sub-dividing the structure into the cylindrical slices. For the *E*-plane waveguide filter, slices would consist of the empty and bifurcated waveguide sections. According to boundary conditions, the complex amplitudes of the modes are matched across the boundary of each slice to the next one; so a set of the linear equations can be derived by matching tangential fields at interfaces and also by solving the simultaneous equations the propagation coefficient and distributions eigenmodes can be obtained. Transverse electric and magnetic modes would be set up, depending on the usage of dielectric in the RF structure. By knowing distributions model coupling integrals between a pair of modes, one each from either sides, can be evaluated at a junction between two cylindrical sections. Hence, complex coupling coefficients which have been achieved are cast into a matrix form. Manufactured filter which is designed based on the mode matching method would not require any further manual tuning, however dimensional uncertainties introduced by manufacturing process must be accurately maintained and controlled for the specified microwave filter system design and analysis [10, 11].

## 2.2. Method 2: Coupled Resonators Technique

Coupled resonator circuits are crucial for design of modern filters, particularly narrow-band and band-pass filters that a play a significant role in various communications applications. This method is based on a general technique for designing coupled resonator filters and is applicable for any type of resonator regardless of its physical structure [11]. Figure 2 presents the block diagram and structure of this design method. This design method is based on coupling coefficients of inter-coupled resonators ($K_c$) and also the external quality factors of the input and output resonators ($Q_{ea}$, $Q_{eb}$). After determining the microwave coupling matrix for the desired filter characteristics, the relationship between the RF values of each coupling coefficient and the structure of coupled resonators would be established to find physical dimensions for fabrication. Coupling coefficient of the RF coupled resonators, regardless of their structure may be defined on the basis of the ratio of coupled energy to stored energy. The filters designed based on this method, have found applications in communication systems. Some of those synthesis methods do not always converge with others such as techniques based on equivalent circuits [12], which provide approximation of the filter parameters. Consequently it is needed to apply optimization methods to meet electrical specifications for RF design [12]. Coupling coefficient ($K_c$) of coupled resonators, regardless of their structure, may be defined on the basis of ratio of coupled energy to stored energy, as in (1); where $E$ and $H$ represent the electric and magnetic field vectors [11].

$$k = \frac{\iiint \varepsilon \underline{E}_1 . \underline{E}_2 \, dv}{\sqrt{\iiint \varepsilon |E_1|^2 \, dv \times \iiint \varepsilon |E_2|^2 \, dv}} + \frac{\iiint \mu \underline{H}_1 . \underline{H}_2 \, dv}{\sqrt{\iiint \mu |H_1|^2 \, dv \times \iiint \mu |H_2|^2 \, dv}}. \tag{1}$$

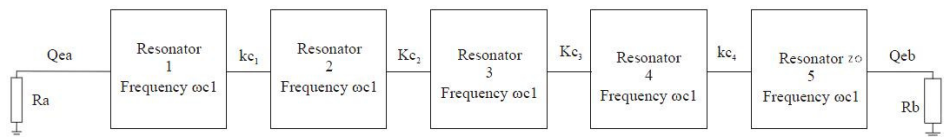

Figure 2. Block diagram of the RF/Microwave filter coupled resonators.

## 2.3. RF Filter Synthesis Formulation Analysis

The microwave and MMW filters transmission and reflection coefficients, $S_{21}$ and $S_{11}$, describe the response of a two-port network. Equations for $S_{21}$ and $S_{11}$ formulated in terms of the coupling matrix may be used for the analysis, synthesis, and design of these resonator RF filters. Coupling coefficients are arranged within a microwave coupling matrix whose inverse is used in order to express the insertion and reflection losses of the structure. The equations for an $n^{th}$ order microwave and MMW filters are given as equation (2) and equation (3) as follows [13]:

$$\xrightarrow{yields} S_{11}(s) = \pm \left(1 - \frac{2}{q_{e1}}[A]_{11}^{-1}\right); \tag{2}$$

$$\xrightarrow{yields} S_{21}(s) = \frac{2}{\sqrt{q_{e1}q_{en}}}[A]_{n1}^{-1}. \tag{3}$$

where the coupling matrix [$m$], forms part of the definition of [$A$] and is given as (4) and (5):

$$\xrightarrow{yields} [A] = \begin{bmatrix} \frac{1}{q_{e1}} & \cdots & 0 & 0 \\ 0 & & 0 & 0 \\ \vdots & \ddots & & \vdots \\ 0 & \cdots & 0 & \frac{1}{q_{en}} \end{bmatrix} + s \begin{bmatrix} 1 & \cdots & 0 & 0 \\ 0 & & 0 & 0 \\ \vdots & \ddots & & \vdots \\ 0 & \cdots & 0 & 1 \end{bmatrix} - j \begin{bmatrix} m_{11} & \cdots & m_{1(n-1)} & m_{1n} \\ m_{21} & & m_{2(n-1)} & m_{2n} \\ \vdots & \ddots & & \vdots \\ m_{n1} & \cdots & m_{n(n-1)} & m_{nn} \end{bmatrix}; \quad (4)$$

$$\xrightarrow{yields} [A] = [q] + s[I] - j[m]. \quad (5)$$

where the variables $q_{e1}$ and $q_{en}$ are the filter external quality factors; [**q**] is an $n \times n$ matrix with all entries zero except for $q_{11} = 1 / Q_{in}$ and $q_{nn} = 1 / Q_{out}$; [**I**] is the $n \times n$ identity matrix; $s = j\omega$ is the filter prototype complex frequency variable; and [**m**] is the RF filter coupling matrix, and its elements (i.e. $m_{ij}$) are known as the coupling coefficients (by varying these values and also the values of the external quality factors, the RF and microwave filter response can be changed). The reflection function $S_{11}$, and transmission function $S_{21}$, can be alternatively expressed as the rational quasi-elliptic ratios of two polynomials, as the following microwave filter equation (6):

$$\xrightarrow{yields} S_{11}(s) = \frac{F(s)}{E(s)}, \qquad S_{21}(s) = \frac{P(s)}{\varepsilon E(s)}. \quad (6)$$

where $F(s)$, $E(s)$, and $P(s)$ are known as the characteristic polynomials; the roots of $P$ and $F$ correspond to the filter's transmission zeros and reflection zeros, respectively; $\varepsilon$ is the ripple constant employed in order to normalize $S_{21}$ to the equiripple level at $s = \pm j\omega$; and the poles common to $S_{11}$ and $S_{21}$ correspond to the roots of the filter's $E(s)$ [13]. The equations in (6) can be directly related to (2) and (3) by using Cramer's rule in order to find the inverse of matrix [**A**] defined in (4) and (5); the inverse relation can be thus given as the equation (7) as follows [13]:

$$\xrightarrow{yields} [A]^{-1} = \frac{adj(A)}{\Delta}. \quad (7)$$

where $\Delta$ and $adj(A)$ are the determinant and adjugate of the filter matrix [**A**], respectively; when $s$ is kept variable, the matrix [**A**] determinant takes the form of a characteristic polynomial with roots as the complex frequency locations of the filtering function poles (i.e. the eigenvalues of matrix [**A**]). It is possible to model the matrix [**A**] characteristic polynomial in order to express the determinant of the filter matrix in terms of its eigenvalues by the filter equation (8) as follows [2]:

$$\xrightarrow{yields} \Delta(s) = s^N - \left(\sum_{i=1}^{N} \lambda_i\right) s^{N-1} + \left(\sum_{i=1}^{N-1} \sum_{k=i+1}^{N} \lambda_i \lambda_k\right) s^{N-2}$$
$$- \left(\sum_{i=1}^{N-2} \sum_{k=i+1}^{N-1} \sum_{m=i+2}^{N} \lambda_i \lambda_k \lambda_m\right) s^{N-3} + \cdots + (-1)^N \prod_{i=1}^{N} \lambda_i. \quad (8)$$

where $N$ is the order of matrix (i.e. filter order), and $\lambda_i$ is the $i^{th}$ eigenvalue of matrix [**A**]; the last term of (8) above is not a departure from the RF pattern observed in the first few coefficients presented. The shift from summation processes to the product-of-series is possible for the last coefficient allowing for a generic expression of the determinant [3]. The mathematical relations in (6) can alternatively be expressed as the following microwave filter equation (9) as well:

$$F(s) = \det([A(s)]) - \frac{2 cof_{11}([A(s)])}{q_{e1}};$$
$$\frac{P(s)}{\varepsilon} = \frac{2 cof_{1N}([A(s)])}{\sqrt{q_{e1} q_{eN}}};$$
$$E(s) = \det([A(s)]). \quad (9)$$

where $det([A(s)])$ presents the filter matrix [**A**] determinant; and $cof_{mn}([A(s)])$ is the matrix [**A**] cofactor. As (9) is not explicitly a polynomial form, an eigenvalue method is used in order to find the roots (i.e. the critical RF frequency) [3]. The equation (5) can be rewritten as (10) as follows:

$$\xrightarrow{yields} [A] = s[I] - (j[m] - [q]) = s[I] - [M']. \tag{10}$$

from the equations (9) and (10), it can be seen that the roots of $E(s)$ are the eigenvalues of the matrix $[M']$; the roots of $P(s)$ are also considered as the generalized eigenvalues of matrix $[M'']$:

$$\xrightarrow{yields} det(s[I'] - [M'']) = 0. \tag{11}$$

where $[M'']$ and $[I']$ have been obtained by deleting the first row and last column of matrix $[M']$ and $[I]$, respectively; hence, the roots of $F(s)$ can be determined by analyzing the $cof_{11}[A(s)]$ and $A(s)$. In the following sections, the design, analysis and synthesis of the four-pole and eight-pole $X$-band and 4-pole $Y$-band RF filters have been comprehensively presented, respectively.

## 3. Microwave and MMW Waveguide Filters: Design and Simulation

### 3.1. 4-Pole Microwave Filter: Method 1

The RF design procedure has been carried out for the design and simulation of the microwave filter at a center frequency of 10 GHz, bandwidth of 500 MHz, maximum $S_{11}$ of –20 dB, and $L_{AR}$ of 0.04321. This design is based on software from the Guided Wave Technology [10] in which the RF dimensions can be achieved by entering the specifications of the filter. The optimization process has also been carried out based on tuning the dimensions of resonators and septa, checking the $S$-parameters of regarding dimensions, and extracting the best result according to the specified characteristics of the microwave filter. Parameter sweep optimization method has been employed in order to accurately optimize the design. Length of resonators and septa after the optimization process can be seen in table 1. As figure 3 shows, the center frequency and bandwidth of the filter is around 10 GHz and 700 MHz, respectively. The number of reflection zeros are 3 and the maximum amount of filter $S_{11}$ has decreased to –17.45 dB.

Table 1. Obtained Dimension Values for the 4-Pole Band-Pass RF Filter Based on Method 1

| | |
|---|---|
| Waveguide Width | 22.86 mm |
| Filter Septum Thickness | 1.0 mm |
| Filter Septum Offset | 11.43 mm |
| Filter Septum Length (1) | 0.28 mm |
| Filter Resonator Length (1) | 14.833 mm |
| Filter Septum Length (2) | 4.83 mm |
| Filter Resonator Length (2) | 15.22 mm |
| Filter Septum Length (3) | 6.112 mm |
| Filter Resonator Length (3) | 15.22 mm |
| Filter Septum Length (4) | 4.83 mm |
| Filter Resonator Length (4) | 14.833 mm |
| Filter Septum Length (5) | 0.28 mm |

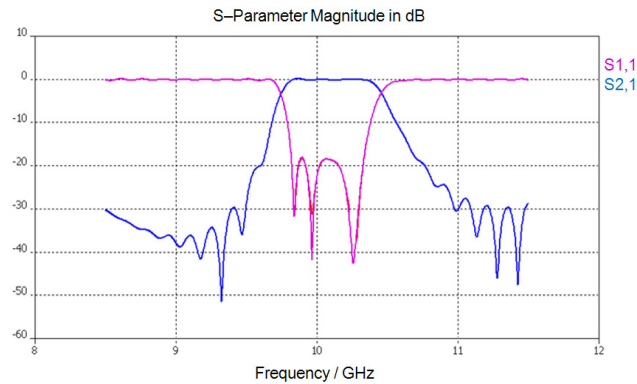

Figure 3. Optimized $S$-parameters of the $X$-band 4-pole filter based on mode matching.

### 3.2. 4-Pole Microwave Filter: Method 2

The synthesis and analysis of the 4-pole microwave filter have been carried out based on the method of coupled resonators. The filter has been designed to have a center operating frequency of 10 GHz and bandwidth of 500 MHz, maximum $S_{11}$ of –20 dB, and $L_{AR}$ of 0.04321. The filter design procedure has been conducted from the ground up and all the design steps are performed and simulated. Appropriate values of the RF coupling coefficients ($K_{ck}$) between resonators and external $Q$ value for the filter input ($Q_{ea}$) and output ($Q_{eb}$) for the corresponding RF filter have been determined. Element values for the Chebyshev low-pass filter prototype filters ($g'_0 = 1.0$, $\Omega_c = 1.0$) for a pass-band ripple $L_{AR} = 0.04321$ dB have also been determined. The low-pass circuit has been modified in order to have a new cut-off frequency $\omega_{c1}$ and the impedance has been transformed to new microwave filter impedance $Z_0$; as shown in table 2.

First step in the RF method is to design a resonator between the two ports that are very weakly coupled. The resonator should resonate at center frequency of the filter, and in order to provide the weak coupling, the maximum amount of $S_{11}$ should be less than –20 dB. In this specific type of RF filter, the resonator is designed by making a gap between two septa which play the role of the resonator and coupling parts, respectively. Varying length of the gap and septa lead to shift in the resonant frequency and cause the maximum value of $S_{11}$, respectively. The required amounts to satisfy the specifications lead to 15.47 mm and 22.926 mm for length of the RF resonator and septum respectively. The next important design step of the method is to establish the relation between the external quality factor of inter-coupled resonators and the filter physical structure of coupled resonators to find length of each external quality factor ($Q_e$). The filter external quality factor controlled by varying the strength of the input coupling of the resonator with a weak output coupling, while all the filter dimensions except the input coupling are fixed. Tuning the length of the input coupling of the resonator leads to different values of quality factor $Q$ in the $S$-parameters of the resonator. The required value of $Q_e$ for the RF filter is calculated as 18.628. The appropriate length in order to have this amount of external coupling for the microwave filter therefore has the value of $l = 0.642$ mm. This accurate design procedure results in the microwave waveguide system structures as follows. The relation between value of coupling coefficients of inter-coupled resonators and physical structure of coupled resonators has been investigated to find the specific length of each coupling. The two resonant peaks that correspond to the characteristics frequencies $f_{p1}$ and $f_{p2}$ are clearly identified from the magnitude responses. The coupling coefficient of the two RF resonators can be determined as (12):

$$\xrightarrow{yields} k = \pm \frac{f_{p1}^2 - f_{p2}^2}{f_{p1}^2 + f_{p2}^2} \qquad (12)$$

Table 2. Determined Values of Coupling Coefficients and External I/O Quality Factor for the 4-Pole Band-Pass RF Filter

| $Q_{ea}$ | $Q_{eb}$ | $K_{c1}$ | $K_{c2}$ | $K_{c3}$ |
|---|---|---|---|---|
| 18.628 | 18.628 | 0.046 | 0.035 | 0.046 |

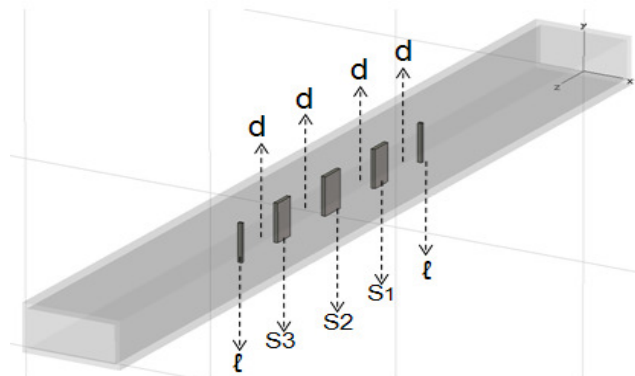

Figure 4. Layout of the *X*-band 4-pole waveguide filter based on coupled resonators.

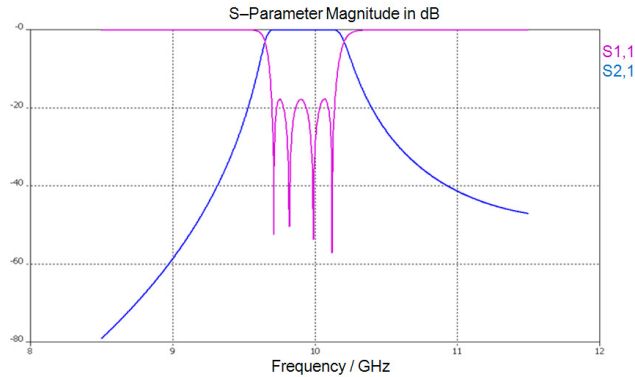

Figure 5. Optimized *S*-parameters of the *X*-band filter based on coupled resonators.

The stronger the coupling, the wider the separation of the two resonant peaks and the deeper the trough in the middle as have been observed from magnitude response [11]. By tuning the coupling between the two coupled resonators in the filter device structure, different values of *K* have been determined. Hence, the physical structure of the *X*-band 4-pole filter is thoroughly known in this stage as the length of all different parts have been found. The length of resonators and coupling parts and the layout along with all the designed internal components of the waveguide filter are shown in detail in the following table 3 and figure 4, respectively.

As figure 5 shows, the microwave filter has been improved in terms of the *S*-parameters based on the optimization of tuning the dimensions of resonators and couplings. The maximum value of $S_{11}$ has decreased to −17.7 dB, the reflection zeros are very deep and the number of them have increased up to the number of the microwave waveguide filter poles.

Table 3. Dimensions of the 4-Pole Microwave Band-Pass Waveguide Filter (Method 2)

| *d* | *l* | $S_1$ | $S_2$ | $S_3$ |
|---|---|---|---|---|
| 15.47 mm | 0.642 mm | 5.044 mm | 6.274 mm | 5.044 mm |

### 3.3. 8-Pole Microwave Filter: Method 2

The microwave waveguide filter design steps include analyzing and extracting the RF resonator size, external quality factor, coupling coefficient (*K*), and further simulation of the filter. The determined RF parameters for the corresponding 8-pole filter including values of coupling coefficients between resonators and also external *Q* values for the input ($Q_{ea}$) and output ($Q_{eb}$), and the dimensions for the microwave waveguide filter are all shown in the following table 4.

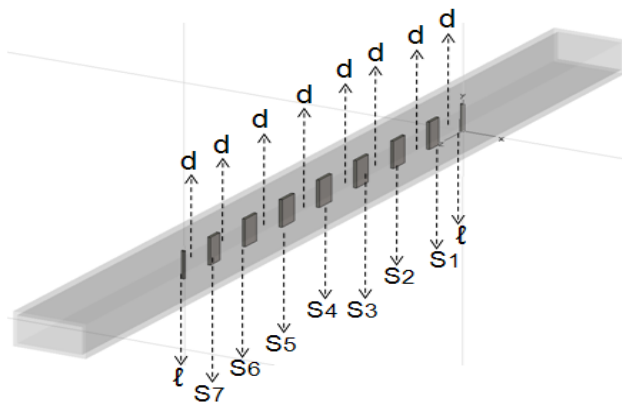

Figure 6. Layout of the *X*-band 4-pole waveguide filter based on coupled resonators.

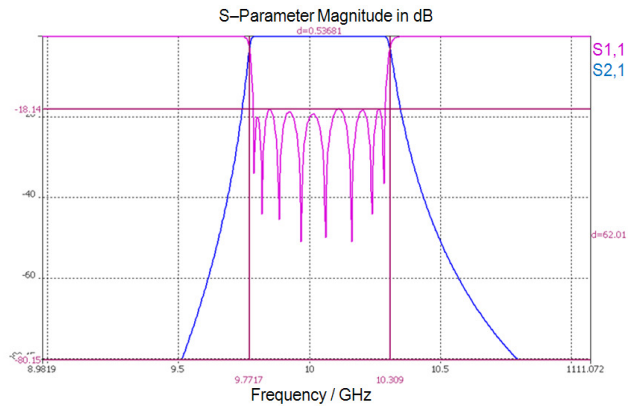

Figure 7. Optimized *S*-parameters of the *X*-band filter based on coupled resonators.

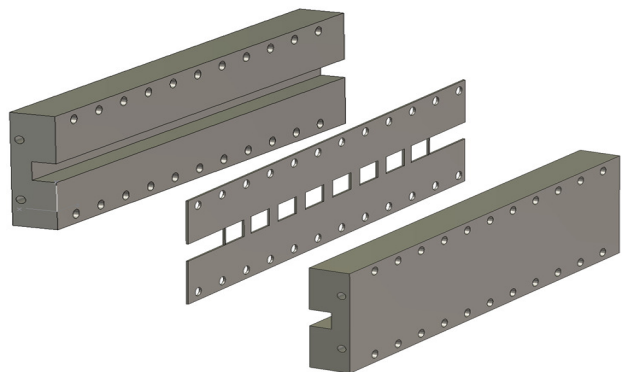

Figure 8. The simulated *X*-band 8-pole microwave waveguide filter based on the novel RF layered structure technology.

Table 4. Dimensions, Coupling Coefficients, and Quality Factors

| $Q_{ea}$ | $Q_{eb}$ | $K_{c1}$ | $K_{c2}$ | $K_{c3}$ | $K_{c4}$ | $K_{c5}$ | $K_{c6}$ | $K_{c7}$ |
|---|---|---|---|---|---|---|---|---|
| 20.3 | 20.3 | 0.04 | 0.03 | 0.028 | 0.027 | 0.028 | 0.03 | 0.041 |

| $D$ | $l$ | $S_1$ | $S_2$ | $S_3$ | $S_4$ | $S_5$ | $S_6$ | $S_7$ |
|---|---|---|---|---|---|---|---|---|
| 15.47 | 0.764 | 5.57 | 7.07 | 7.39 | 7.55 | 7.39 | 7.0 | 5.57 |

Figure 6 indicates the layout of the filter as a result of the accurate 8-pole microwave filter analysis and design based on the coupled resonators. Figure 7 shows the *S*-parameters of the filter and optimization based on tuning the length of different sections to achieve better filter performance. It can be seen from figure 7 that the center frequency is equal to 9.9 GHz with the RF bandwidth of 500 MHz, the number of filter reflection zeros is equal to the number of poles, and the maximum amount of filter $S_{11}$ is equal to −18.14 dB. Figure 8 also presents this novel RF layered technology introduced for the waveguide filters (both the RF 4-pole and RF 8-pole).

### 3.4. 4-Pole Millimeter-Wave Filter: Method 2

The appropriate millimeter-wave (MMW) rectangular waveguide for the corresponding filter has been designed. Figure 9 presents the simulated *S*-parameters. The value of $S_{21}$ for the corresponding rectangular waveguide is equal to zero for the RF frequency range of 220-325 GHz, which indicates complete transition of signal from port 1 to port 2 of the waveguide. The waveguide has the cut-off frequency of 196.71 GHz in the specified range. Table 5 and figure 10 show the 4-pole millimeter-wave filter determined design parameters and obtained layout, respectively. Figure 11 illustrates the analyzed and computed *S*-parameters for the filter. As it can be seen, the center frequency of the filter is around 300 GHz and the bandwidth is around 10 GHz. Four deep reflection zeroes has appeared in this design and the maximum value if $S_{11}$ is around –10 dB which needs to be further enhanced and optimized to improve the results.

Table 5. 4-Pole Band-Pass Millimeter-Wave Filter Dimensions and Determined Values of Coupling Coefficients and External I/O Quality Factor (Coupled Resonators)

| $Q_{ea}$ | $Q_{eb}$ | $K_{c1}$ | $K_{c2}$ | $K_{c3}$ |
|---|---|---|---|---|
| 46.57 | 46.57 | 0.018 | 0.014 | 0.018 |

| $D$ | $l$ | $S_1$ | $S_2$ | $S_3$ |
|---|---|---|---|---|
| 0.3494 mm | 0.6222 mm | 1.889 mm | 1.926 mm | 1.889 mm |

## 4. Microwave Waveguide Filters: Realization and Measurement

The structure of the rectangular waveguide in X-band is based on three layers and is made of aluminum. The fabrication (figure 13) was carried out and the rectangular waveguide was extensively measured on a network analyzer. The measured amount of waveguide $S_{21}$ in X-band is equal to zero to allow the complete transition of from port 1 to port 2 of the rectangular waveguide. After the fabrication, the regarding filter has been tested by the network analyzer. Figure 13 presents this fabricated X-band filter and measured S-parameters, respectively. It is indicated, the regarding microwave layered filter has the center frequency and bandwidth of 9.9 GHz and 700 MHz, respectively. The number of filter reflection zeros are four and the maximum amount of $S_{11}$ is around –19.3 dB. The filter experimental results agree with the simulations.

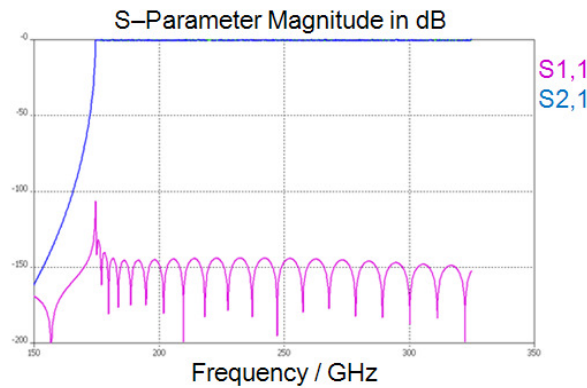

Figure 9. Computed S-parameters of the Y-band rectangular MMW waveguide (WR3).

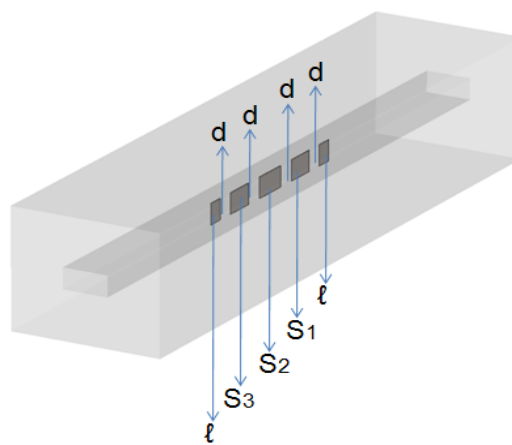

Figure 10. Layout of the designed and simulated Y-band 4-pole MMW waveguide filter based on coupled resonators (microwave filter design; method 2).

The fabrication process has also been carried out for 8-pole filter and the filter has been extensively tested on a network analyzer. Figure 14 shows the measured parameters of 8-pole filter which has the center frequency of around 10 GHz and filter bandwidth of approximately 500 MHz; the number of filter reflection zeros is seven, and the maximum value of waveguide filter $S_{11}$ is around –15.7 dB with the novel introduced three-layer RF and microwave structure.

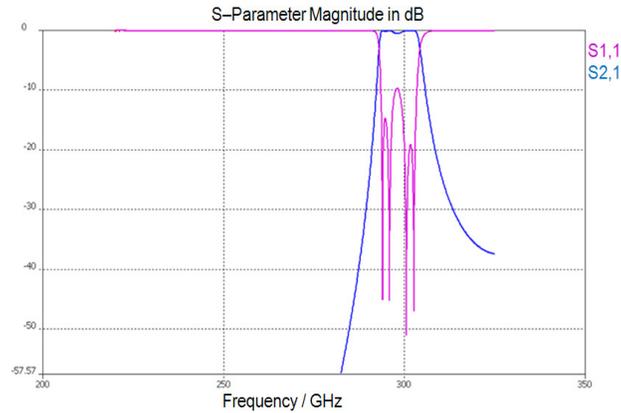

Figure 11. *S*-parameters of the MMW waveguide filter based on coupled resonators.

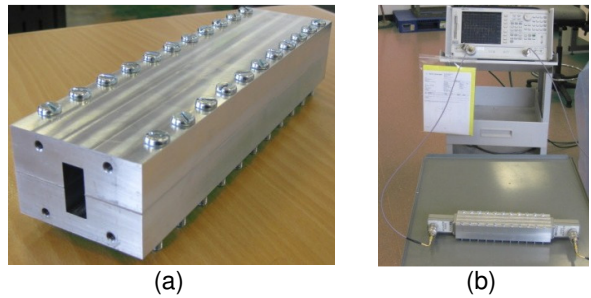

(a)          (b)

Figure 12. Filter configurations: (a) fabricated RF filters layered structure; (b) filter microwave measurement set up.

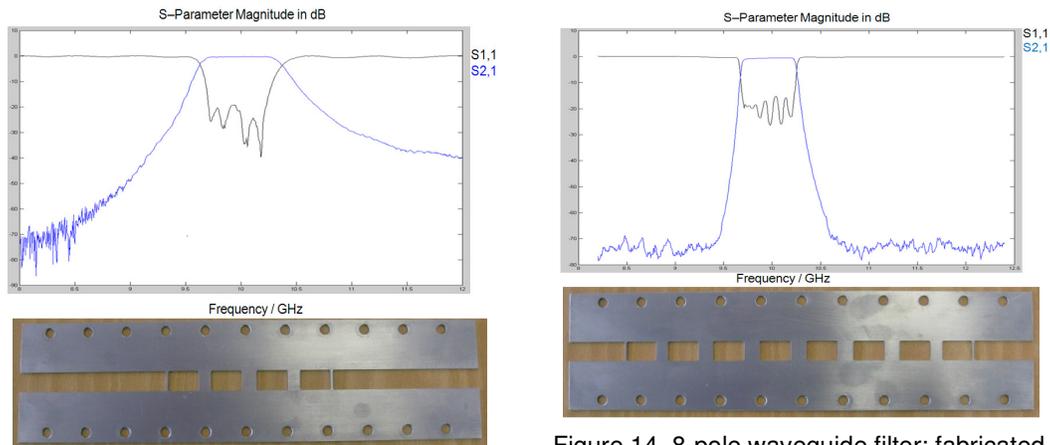

Figure 13. 4-pole waveguide filter: fabricated internal part; and the measured *S*-parameters.

Figure 14. 8-pole waveguide filter: fabricated internal part; and the measured *S*-parameters.

## 5. Conclusion

In this contribution, several promising designs for the microwave and millimeter-wave *E*-plane filters have been comprehensively carried out for operation at *X*-band and *Y*-band. The waveguide filters have been accurately designed, synthesized, analyzed, fabricated, and tested based on the mode matching and coupled resonators techniques and the output microwave characteristics have been shown and analyzed in terms of the RF filters obtained *S*-parameters. The designs provide waveguide filters with significant output results. These filters can further be optimized using evolutionary computation and also numerical electromagnetic methods. For the fabricated filter structures, there are some screws needed to be located within cavities of the filters so that by tuning these screws, dimensions of the resonators gets affected, hence the improved *S*-parameters can further be obtained as a result of the design process. The *Y*-band filter can be fabricated using gold; it should be noted that this particular filter has tiny dimensions and the measured results will have different *S*-parameters in compare with the RF simulations.

**References**
[1] Shang X, Ke M, Wang Y, Lancaster MJ. Micromachined W-band waveguide and filter with two embedded H-plane bends. *IET Microwaves, Antennas & Propagation*. 2011; 5(3): 334-339.
[2] Jayyousi AB, Lancaster MJ. Analytic sensitivity calculations of poles and zeros of general Chebyshev filtering functions for microwave coupled resonators. *IET Microwaves, Antennas & Propagation*. 2010; 4(7): 893-898.
[3] Shang X, Wang Y, Nicholson GL, Lancaster MJ. Design of multiple-passband filters using coupling matrix optimisation. *IET Microwaves, Antennas & Propagation*. 2012; 6(1): 24-30.
[4] Wang X, Bao P, Jackson TJ, Lancaster MJ. Tunable filters based on discrete ferroelectric and semiconductor varactors. *IET Microwaves, Antennas & Propagation*. 2011; 5(7): 776-782.
[5] Shang X, Lancaster MJ, Ke M, Wang Y. Measurements of micromachined submillimeter waveguide circuits. In *Proceedings of the 76$^{th}$ Microwave Measurement Symposium*. 2010.
[6] Bui LQ, Ball D, Itoh T. Broadband millimeter-wave E-plane bandpass filters. In *Proceedings of the IEEE MTT-S International Microwave Symposium Digest*. 1984.
[7] Postoyalko V, Budimir DS. Design of waveguide E-plane filters with all-metal inserts by equal ripple optimization. *IEEE Transactions on Microwave Theory and Techniques*. 1994; 42(2): 217-222.
[8] Cronin NJ. *Microwave & Optical Waveguides*. Taylor & Francis, 1995.
[9] Ruiz-Cruz JA, Zak, KA, Montejo-Garai JR, Rebollar JM. Rectangular waveguide elliptic filters with capacitive and inductive losses irises and integrated coaxial excitation. In *Proceedings of the IEEE MTT-S Int. Microwave Symposium Digest*. 2005.
[10] Guided Wave Technology Online Microwave Filter Design. (2012). [Online]. Available at: http://www.guidedwavetech.com/.
[11] Hong JS, Lancaster MJ. *Microstrip Filters for RF/Microwave Applications*. 1$^{st}$ ed. John Wiley. 2001.
[12] Yu M, Wang Y. Synthesis and beyond. *IEEE Microwave Magazine*. 2011; 12(6): 62-76.
[13] Nicholson GL, Lancaster MJ. Coupling matrix synthesis of cross-coupled microwave filters using a hybrid optimisation algorithm. *IET Microwaves Antennas & Propagation*. 2009; 3(6): 950-958.